\documentstyle[prb,aps]{revtex} 
\input psfig.tex
\newcommand{\beq}{\begin{equation}}
\newcommand{\eeq}{\end{equation}}

\newcommand{\z}{\zeta}
\begin{document}
\draft
\twocolumn[\hsize\textwidth\columnwidth\hsize\csname @twocolumnfalse\endcsname

\title{Ostwald Ripening in Two Dimensions: Treatment with Pairwise Interactions.}
\author{Boris Levitan and Eytan Domany}
\address{Department of Physics of Complex Systems, Weizmann Institute
of Science, Rehovot, Israel}
%\date{\today}
\date{July 7, 1997}
\maketitle
\begin{abstract}
We present a systematic extension of mean field models for Ostwald ripening in
two dimensions. 
We derived, using a mean field type approximation, an analytic expression
for pairwise interactions between the minority phase droplets. 
These interactions appear in  dynamic equations for 
the droplets' radii and positions, which 
are solved numerically for systems
of tens of thousands of droplets, yielding very good agreement with recent
experiments.
\end{abstract}
\pacs{PACS numbers: 64:60My, 64.60Cn, 64.75+g}
\vskip2pc]
Ostwald ripening\cite{Ostwald} is a coarsening process, observed during the
late stage of the evolution of a two-phase system (say, solid in liquid), in
the course of which the droplets of the minority phase exchange material by
means of diffusion.  This process leads towards a scaling state in which the
characteristic length scale grows with time according to the Lifshitz-Slyozov
law\cite{LifshSlyoz} $\bar R(t)\sim t^{1/3}$. When rescaled by $\bar R(t)$, all
statistical characteristics of the system (such as droplet size distribution,
spatial correlations etc) are time-independent.

These characteristics were studied in a number of detailed numerical
simulations. The Cahn-Hilliard equation\cite{RogersDes,Masbaum}, solved in the
entire system, provides the most  basic model of Ostwald ripening. If the solid
droplets are assumed to be uniform and only their boundaries are retained to
describe the system, the problem becomes one of solving the diffusion equation
for the concentration field $c(\vec r)$ between droplets, with Gibbs-Thomson
boundary conditions\cite{note1}:
%note1} $R(t)$ is the local radius of curvature. 
\beq
c|_{droplet}(t)=c_{eq}(R(t))=c_{\infty}+{\alpha\over R(t)}.
\label{eq:GT}\eeq
The boundaries  move in response to the incident flux; 
\beq
\frac{d R}{d t} = -J_\perp  \qquad \qquad \vec{J} = -\ \nabla c|_R
\label{eq:J}
\eeq 
At long times a quasi-static approximation can be used and the diffusion 
problem turns into Poisson's equation in 2-$d$,
\beq
\nabla^2 c= 0.
\label{eq:Poiss}
\eeq
This problem can be reduced to an implicit system of ordinary differential
equations\cite{Kawasaki,AkaiwaV,AkaiwaM}. Since these calculations take into
account $all$ the complicated interactions between the droplets, mediated by
the diffusion field, they do not elucidate the relative importance of different
aspects of these interactions. The number of droplets and the length of the
simulations are also limited. On the other end of the complexity scale are
analytical mean field treatments \cite{LifshSlyoz,Wagner,Marq} that neglect
$all$ spatial effects; these, however, are too simple to account for
correlations in a system of non-vanishing volume fraction.  Previous attempts
to "interpolate" between these extremes, and provide an analytical treatment of
the spatial effects\cite{Marder,ZhGunt} contain uncontrolled approximations and
lead to very complicated expressions.

Using a mean-field type approximation, we calculated {\it analytically} a
pairwise interactions between the droplets. These appear in closed-form
dynamic equations for many interacting droplets which are integrated
numerically. 
Our model gives rise to 
a very efficient numerical algorithm; the evolution of tens of thousands of
droplets can be followed.
Our work was motivated by and our
results are compared with a recent experiment on a two-dimensional film of
liquid and crystalline succinonitrile in coexistence\cite{KS}. This experiment
shows that even at $\varphi =0.4$ the droplets are almost circular; therefore
we characterize the system by the set of the droplets' radii $R_i$ and the
positions $\vec r_i$ of their centers.  

We write the solution of Eq.(\ref{eq:Poiss}) in terms of "charges" and 
"dipoles" 
placed at $\vec r_i$      
\beq
c(\vec r)= \sum_i q_i  \log ( |\vec{r} - \vec{r}_i| / R_0 )+
\sum_i {(\vec p_i\cdot(\vec{r}-\vec r_i))\over |\vec{r} - \vec{r}_i|^2}
\label{eq:sol}\eeq
($R_0$ is an arbitrary length). The boundary conditions (\ref{eq:GT}) on the
surface of each droplet give rise to linear equations which determine the
charges $q_i$ and the dipoles $\vec p_i$. When the resulting solution is used in
(\ref{eq:J}) to derive the flux at droplet $i$, we find that it's normal
component has two terms. One is isotropic, due to the charges; it affects the
droplet's area and therefore determines the dynamics of the radii. The other is
an anisotropic contribution, due to the dipoles, which gives rise to a shift of
the centers of the droplets\cite{Marder,Long}. The charges and dipoles can be
eliminated, yielding\cite{Long} dynamic equations for the radii and positions
of the droplets: 
\beq
\dot R_i=\frac{q_i}{R_i}=\sum_j L_{i,j}^{-1}(1+R_j/R_c)=\sum_j L_{i,j}^{-1},
\label{dRdt}  
\eeq
\beq
{d\vec r_i\over dt} =-{2\vec p_i\over R_i^2}=
2\sum_{j\ne i} R_j {\dot R_j}
%{q_j\over |X_{i,j}|}
{\vec X_{i,j}\over |X_{i,j}|^2},
\label{shift}
\eeq
where $X_{i,j}=|\vec r_j-\vec r_i|$ and $R_c=\alpha/c_\infty$ is a capillary 
length. The second equality in Eq.(\ref{dRdt}) is due to a sum 
rule\cite{Long,BeenRossI}; $L_{i,j}^{-1}$ are the elements of the $inverse$ of 
the matrix $\hat L$, defined as\cite{Kawasaki}:
\beq
L_{i,j} =R_iR_j\cases{\log (R_i/R_0) & if $i=j$\cr
			   \log (X_{i,j}/R_0) & otherwise\cr}
\label{eq:Lij}\eeq
The value of the parameter $R_0$ is $arbitrary$; it does not affect the dynamics
of the system\cite{Long}. Once $\hat L$ has been inverted, we can evaluate the
radii at the next time step. One must invert the $N\times N$ matrix $L_{i,j}$
at $each$ time step (which takes $O(N^3)$ operations); since we need O(N) time
steps, each run would cost $O(N^4)$ computations.  

Beenakker\cite{Been} solved an analogous problem in 3-$d$ by $truncating$ the
matrix $L_{i,j}$. Motivated by the effect of {\it
screening}\cite{MarderI,MarqRoss}, he took into account only interactions
between those droplets whose separation does not exceed a threshold. Akaiwa and
Meiron\cite{AkaiwaM} used the analogous truncation procedure in 2-$d$. However,
formal truncation of the matrix seems problematic. Since the matrix elements
$grow$ with $X_{i,j}$, the elements of the $inverse$ matrix $\hat L^{-1}$ as
functions of the cutoff should contain large fast oscillating components. 
Apparently, the success of Akaiwa and Meiron\cite{AkaiwaM} is due to the fact
that these oscillations are effectively averaged out during the run.  

Rather than discarding large matrix elements 
and proceeding to invert the truncated matrix numerically, we calculate 
analytically the elements of
$\hat L^{-1}$, using a mean field approximation. 
First, represent $\hat L$ as
the sum of its diagonal and off-diagonal parts,
${\hat L}=
{\hat L}_0 - {\hat L}_1$,
where $(\hat L_0)_{i,j}=\delta_{i,j}R_i^2\log (R_i/R_0)$ 
and $(\hat L_1)_{i,j}=-(1-\delta_{i,j})R_iR_j\log(X_{i,j}/R_0)$
and introduce a new matrix $\hat T$ defined by 
$$\hat L^{-1}=\hat L_0^{-1} +\hat L_0^{-1}\hat T \hat L_0^{-1}$$ 
This is indeed the inverse of $\hat L$ if
$\hat T$ satisfies the equation
\beq 
\hat T=\hat L_1+\hat L_1\hat L_0^{-1}\hat T
\label{T}
\eeq
For convenience we represent $T_{i,j}=R_iR_j\phi_{i,j}$; Eq.(\ref{T}) now
becomes a condition on the matrix $\hat \phi$, which we write separately for
the diagonal and off-diagonal elements:
\beq
\phi_{k,k}= 
-\sum_{j \neq k}\frac{\log (X_{k,j}/ R_0)}{ \log (R_j/R_0)}  \phi_{j,k}
\label{exd}\eeq 
\beq
\sum_{j \neq i,k}\frac{\log (X_{i,j}/ R_0)}{ \log (R_j/R_0)}  \phi_{j,k}+
\phi_{i,k}= -\gamma_k \log (X_{i,k}/ R_0)
\label{ex}\eeq
with 
\beq
\gamma_k=1+{\phi_{k,k}\over\log (R_k/R_0)}.
\label{gamma}\eeq
The set of equations (\ref{exd}-\ref{gamma}) is exact but solving it is as
difficult as inverting $\hat L$. The advantage of this formalism is that it
serves as a convenient starting point to generate {\it approximate} expressions
for $\phi_{i,j}$ - in particular, a manageable mean-field approximation. 

We give now an outline of the main ideas and formal steps of the derivation
of our approximation to $\phi_{i,j}$; details will be presented elsewhere
\cite{Long}. 
First of all we assume that the $i,j$ matrix element depends only on the
corresponding distance\cite{foot}, 
$\phi_{i,j}=\phi(X_{i,j})$, and that $\phi(X)$ 
%is a smooth function. 
varies only on scales $\zeta$ such that the number of droplets within an area
$\zeta^2$ is large. If this holds we can replace in  Eqs.(\ref{ex}) and
(\ref{exd}) the factor $1/\log (R_j/R_0)$ by its mean value and the sums by
integrals. Within this mean-field approximation Eq.(\ref{exd}) implies
$\phi^{(mf)}_{k,k}=$const$.=\phi_0$; this leads to a self-consistent
expression\cite{Long} for $\phi(X_{i,j})$ only if we set $\gamma_k=$
const.$=\gamma_0$ which, in turn, means (see Eq.(\ref{gamma})) that $\phi_0=0$
and $\gamma_0=1$. Thus within our mean-field treatment of $\phi$, Eq.(\ref{ex})
becomes 
\beq
{1\over 2\pi\z^2} \int\log ({X_{i,j}\over R_0})\phi(X_{j,k})d^2r_j -
\phi(X_{i,k})=\log {X_{i,k}\over R_0}
\label{naiv}\eeq
where we introduced a constant length, defined by
\beq  
\z^{-2}=2\pi n\left\langle {1\over\log (R_0/R)}\right\rangle
\approx 2\pi n {1\over\log (R_0/\left\langle R \right\rangle)}
\label{z}\eeq

Here $n= \varphi/\pi\langle R^2\rangle$ is the number of droplets per unit
area; the angular brackets denote averaging over the droplet size distribution.
Operating with $\bigtriangledown_i^2$ on Eq.(\ref{naiv}) 
%
%an using the identity
%$\bigtriangledown^2\log r=2\pi \delta({\vec r})$ 
%we obtain the following
yields the differential equation
\beq  
-\z^{-2}\phi+\bigtriangledown^2\phi=-2\pi\delta({\vec r}),
\label{Bes}\eeq
which is solved by the zeroth order modified Bessel function of the second kind
(also called MacDonald function), $\phi(X)=K_0(X/\z)$. This determines the
$off-diagonal$ elements of $\phi_{i,j}$ up to the length $\z$ which depends on
the arbitrary parameter $R_0$. Since the value of this parameter does not
affect the {\it exact} solution of the problem, we can tune it to improve our
approximation. Note that when our $\phi(X)$ is used in the integral form of 
Eq.(\ref{exd}) we get 
\beq
{1\over 2\pi\z^2}
\int\log (X_{k,j}/R_0)K_0(X_{j,k}/\z)d^2r_j= \phi_{k,k}=\phi_0=0
\label{diag00}\eeq
After some lengthy algebra, 
presented elsewhere\cite{Long}, we can show that this
condition is satisfied when $R_0$ satisfies
the following 
{\it approximate} expression:
\beq
\left\langle {1\over\log (R_0/R)}\right\rangle 
\approx \left\langle {1\over K_0( R /\z)}\right\rangle
\label{con}\eeq
By substituting Eq.(\ref{con}) in Eq.(\ref{z}) we eliminate $R_0$ 
from the problem and obtain an equation for $\z$:
\beq
\z^{-2} = 2\pi n
\left\langle{1\over K_0( R /\z)}\right\rangle,
\label{z0}\eeq
which is almost identical to Marqusee's expression for the screening
length\cite{Marq}. The solution for $\phi(X)$  determines the
matrix $\hat T$ and we get an expression for the matrix elements $L_{i,j}^{-1}$
which, when used in Eq.(\ref{dRdt}) together with the relation\cite{Long}
\[
-\sum_{j \neq i} \frac{1}{\log (R_j/R_0)} K_0({X_{i,j}\over\z}) 
\approx \frac{1}{2 \pi\z^2} \int K_0 ({r\over\z}) d^2r = 1,
\]
finally yields our central result:
\begin{eqnarray} 
R_i{dR_i\over dt} = R_i&\sum_{j}& L^{-1}_{i,j} = \frac{1}{K_0(R_i/\z)}\times
\nonumber \\
&\times&
\sum_{j (\neq i)}
\frac{K_0(X_{i,j}/\z )}{K_0 (R_j /\z)} 
\left( \frac{1}{R_j}-\frac{1}{R_i} \right).
\label{Linv}
\end{eqnarray}
Since $K_0(x)\sim \exp(-x)$ for large $x$, $\z$ indeed has the meaning of a
screening length. Note that the total area of the droplets is conserved by
Eq.(\ref{Linv}).  

The approximations made, of replacing the sums by integrals and
$1/\log(R_j/R_0)$ by its average value $1/(2\pi\z^2)$, are valid if the
function $\phi(X)$ is smooth and the number of  droplets in the screening zone
is large: $N_{\z}=n\z^2\gg 1$. For  $\varphi=0.13$  (as used in the
experiment) the approach outlined above gave a much too small value for $\z$. A
somewhat larger value was obtained\cite{Long} by taking into account the fact
that each of the droplets is surrounded by a {\it depletion zone}, from which
all neighbors are excluded. Including depletion zones retains an equation of
the same form as Eq.(\ref{Linv}) for the evolution of the radii, but with the
mean-field $\z$ replaced by $\z_{sc} > \z$.  For $\varphi=0.13$ it becomes
$\z_{sc}=2.73\bar R\approx 0.56X$ where $X$ is the typical distance between
neighbors. Although $\z_{sc}$ is still too small to provide a formal support
to our approach, we tested it by integrating (numerically) Eqs.(\ref{Linv}) and
(\ref{shift}).

The detailed description of our numerical algorithm is presented
elsewhere\cite{Long}. An advantage of our method is that it conserves  the
total area of the droplets at each time step. Computational efficiency is very
high: an entire run, starting with $N$ droplets initially and running till most
disappear (i.e. reach $R_i \approx 0$) takes only $O(N\log N)$ operations. This
is due to the fact that we are able to keep the time step reasonably large,
eliminating a large number of droplets at each step.  At the same time, unlike 
previous studies \cite{AkaiwaM,Been,YaoI}, we ensure $exact$ conservation of
the total area of the droplets at each time step.  This makes our approach
useful for an extensive study of the Ostwald problem.

We used initial systems of $50 000$ droplets with toroidal boundary conditions.
First evaluate $\dot R_i$ using Eq.(\ref{Linv}) with $\z=2.73\bar R$; next use
Eq.(\ref{shift}) to calculate the motion of the droplets' centers.

Regarding this motion, one can use Eq.(\ref{shift}) to estimate its importance
\cite{Marder,Long} by comparing the characteristic times for significant shift
versus a shrinking droplet's lifetime;
$\tau_{\rm shift}/\tau_{\rm life}\sim {X^2\over R^2}\approx \pi/\varphi$.
Hence for small fractions (and even  for $\varphi=0.13$) the shift of the
droplets is 
\vbox{ 
\begin{figure}
\centerline{\psfig{file=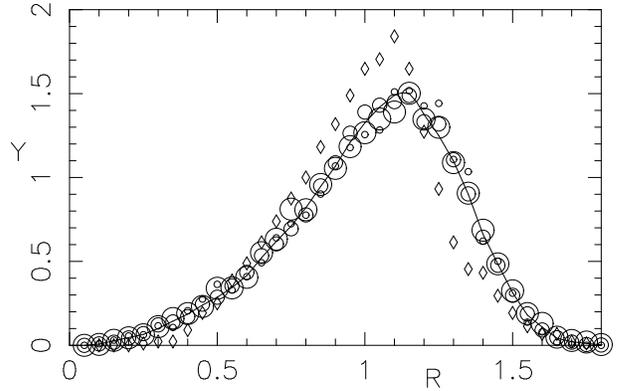,width=3.3 truein}}
% 50 20 650 430  (width=3.0)
\caption{
%fig. 1
The  size distribution of the droplets; $R$ is normalized by $\bar R$. Data
were taken when the number of droplets (initially 50 000) was $N=3000$ ({\it
largest circles}), $N=2000$ ({\it medium circles} and $N=1000$ ({\it small
circles}. Each plot presents the average over 8 runs; the solid line is a
guide to the eye. The fact that at these three times the same distribution 
was obtained indicates that we reached the scaling state. 
The experimental result of Krichevsky and Stavans is also shown ({\it diamonds}).  
}
\end{figure}
}
$adiabatically$ slower than the droplets' growth and one is tempted
to neglect the motion of the centers. Indeed we tried this, but the resulting
solution of the dynamics of the radii alone led to severe inconsistencies
(manifest in significant overlap of neighboring droplets, i.e. $R_i+R_j >
X_{i,j}$). This convinced us that the correlations induced by the proper
movement of the droplets are essential for a physically sensible solution.

The sum on the r.h.s. of Eq.(\ref{shift}) appears to have convergence problems;
assuming that the charges $q_j$ are uncorrelated random variables, with 
zero mean (as required by the total area conservation) one can easily get 
\beq
\left\langle\left({d\vec  r_i\over dt}\right)^2\right\rangle \sim
 \langle q^2\rangle\sum {1\over |X_{i,j}|^2}\sim \langle q^2\rangle\log L_s,
\eeq
where  $L_s$ is the size of the system. The charges, however, {\it are}
correlated (the effect of screening) and this causes fast
convergence\cite{Long}. We performed in some cases the entire sum and then
repeated the simulations, summing only over droplets $j$ with $|X_{i,j}|<b$,
using $b=2.3X$ and $b=3.25X$ (the latter value corresponds to about $10$ terms
in the sum); no significant difference between these three simulations was
found.

For 50,000 initial droplets at $\varphi=0.13$ the system approached the scaling
state at $N\approx 3000$; this is when we started our measurements. In order to
reduce fluctuations we averaged all the data over 8 runs. Such a run took
about 105 minutes on a single processor of an HP-9000 (series K200)
workstation.  First we present the distribution of the droplets' radii in the
scaling state (see Fig. 1). The agreement with the experimental points\cite{KS}
is reasonable.  
\vbox{ 
\begin{figure}
\centerline{\psfig{file=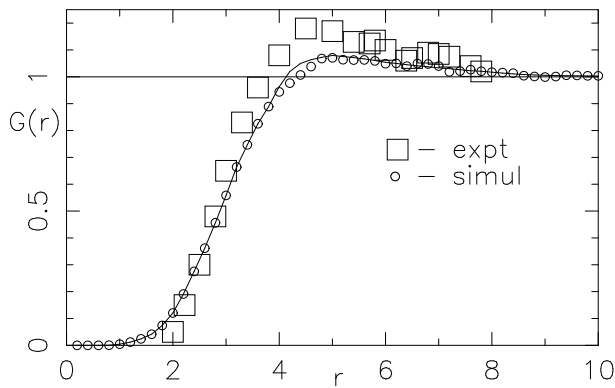,width=3.3 truein}}
\caption{
%fig. 2 
Weighted time average of the correlation functions of droplets' positions,
$G(r)$, taken at different times in the scaling state, 
each averaged over 8 runs 
({\it circles});  the experimental results of Krichevsky and Stavans
({\it squares}). The solid line is a guide to the eye.
}\end{figure}}
A much more demanding comparison is that of various correlation functions 
%that were measured by Krichevsky and Stavans. 
Fig. 2 presents the correlation
function $G(r)$ - the probability to find a droplet's center at a distance $r$
from a given droplet's center, as obtained by our numerical solution and
experiment\cite{KS}. We recover the positions of the initial rise of $G(r)$ as well as
the existence and position of its maximum. The value of our curve at the
maximum is smaller than the experimental one; our $G(r)$ is rather close to the
result of Masbaum, obtained by solving the Cahn-Hilliard equation (see Fig. 1a
in ref.\cite{Masbaum}), that also exhibits a small peak.  

We  measured also the correlation functions of the {\it charges} defined first
in ref.\cite{KS}, which contain more detailed information about the system. 
For a charge $q_i$ calculate $Q_+(r)$, the total amount of similar charge as
$q_i$ within an annulus $[r,r+dr]$ around $\vec r _i$ and define the function
$$g_+(r) = \langle q_i Q_+(r) \rangle.$$
Similarly we define $g_-(r)$ in terms of the {\it opposite} charges. These two
functions, as obtained by our simulations, are presented in Fig. 3
together with the corresponding experimental data of Krichevsky and Stavans. 
The agreement between theory and experiment is quite impressive.

In summary, we introduced and tested a model for the Ostwald ripening process
in two dimensions. Our model approximates the exchange of material between the
droplets, as determined by the complicated diffusive interaction, by simple
pairwise couplings. An approximate mean field type approach is used to evaluate
these pairwise interactions. The resulting dynamic equations are then solved
numerically. Our model gives good agreement with experiment for a fairly large
value of area fraction. We found that the shift
of the droplets plays an important role at $\varphi=0.13$. Our method leads to
a very 
efficient numerical algorithm that can be useful for future studies of
this problem. This research was sup-
\vbox{
\begin{figure}
\centerline{\psfig{file=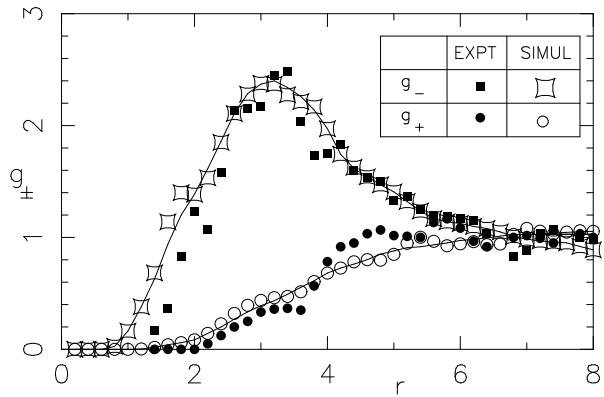,width=3.3 truein}}
\caption{
%fig. 3
The correlation functions for the same and opposite charges, as obtained by our
simulations and by  experiments. Our data present averages over 8 runs. The 
solid line is a guide to the eye.
}
\end{figure}
}
ported by grants from the Germany-Israel Science 
Foundation (GIF). B. L. thanks the Clore Foundation for financial support.
We thank O. Krichevsky, J. Stavans and D. Kandel for discussions.
%References


\begin{references}
\bibitem{note1} 
$R(t)$ is the local radius of curvature.
\bibitem{foot} In general the diffusion field induces "many-body" interactions 
between droplets. Our assumption amounts to keeping pairwise interactions
only.
\end{references}
\end{document}